\documentclass[12pt,a4paper]{article}
\usepackage[latin2]{inputenc}
\usepackage[T1]{fontenc}
\usepackage{graphicx}
\usepackage{subfigure}
\usepackage{amsfonts}
\usepackage{amssymb}
\usepackage[intlimits]{amsmath}
\usepackage[mathscr]{eucal}
\usepackage{indentfirst}
\usepackage{bbm}
\usepackage{verbatim}
\makeatletter
\newcommand{\df}[2]{\frac{\partial #1}{\partial #2}}
\newcommand{\bigx}[1]{\bBigg@{#1}}

\newcommand{\ud}{\mathrm d}

\makeatother

\numberwithin{equation}{section}

\begin{document}

\title{{\normalsize\begin{flushright}ITP Budapest Report
    No. 633 \end{flushright}}\vspace{1cm}Boundary renormalisation group flows of the
    supersymmetric Lee\,--Yang model and its extensions}

\author{M\'arton Kormos\footnote{kormos@general.elte.hu
}\\
\emph{\small Institute for Theoretical Physics, E\"otv\"os University, }\\
\emph{\small 1117 Budapest, P\'azm\'any P\'eter s\'et\'any 1/A, Hungary}\\
}

\date{May 2, 2007}

\maketitle
\thispagestyle{empty}

\begin{abstract}
In this paper we examine the supersymmetric Lee\,--Yang model in the
presence of boundaries. We determine the reflection factors for the
Neveu--Schwarz type boundary conditions from the reduction of the
supersymmetric sine--Gordon model and check them by using Boundary
Truncated Conformal Space Approach in the massless case. We explore
the boundary renormalisation groups flows using boundary TBA and TCSA.
\end{abstract}

\newpage

\begin{section}{Introduction}
Boundary conformal field theories attracted much interest recently,
due to their applications in describing D-branes in string theory
\cite{openstrings,tachyons} and their relevance in condensed matter
physics, e.\ g.\ in the Kondo problem \cite{saleur}. 

Several papers appeared in the literature that deal with the
consistent boundary conditions and reflection factors
\cite{nepomechie,tgzs,gabor,baseilhac,bssgrevisited,ahnsuzuki}, the
boundary perturbations and the corresponding renormalisation flows
\cite{triflows,feverati,an,kormos} of two-dimensional supersymmetric
field theories.

One of the simplest supersymmetric theory is the supersymmetric
Lee\,--Yang (SLY) model and in this paper we study this model in the
presence of boundaries. In a paper \cite{an} Ahn and Nepomechie
proposed reflection factors for the superconformal boundary conditions
and determined a boundary flow using Boundary Thermodynamic Bethe
Ansatz (BTBA). Here we argue that both their reflection factor and one
of their TBA formulae are mistaken. We propose different reflection
factors which we check using Boundary Truncated Conformal Space
Approach (BTCSA). We also predict a different boundary flow using BTBA
and we compare the fixed points and the change of the boundary entropy
again with our BTCSA results.

The paper is organized as follows. In section 2 we summarize some
basic facts about the SLY model. In section 3 we determine the
reflection factors by applying a ``folding'' trick to the
supersymmetric sine--Gordon reflection factors. In section 4 some
details of the BTCSA are explained briefly and after calculating the
massless reflection factors the predicted energy levels and phase
shifts are compared to the BTCSA results. In section 5 we write down
the BTBA equations and calculate the variation of the \(g\)-function
along the boundary flow. In section 6 this is compared to the BTCSA
result for the fixed points and the \(g\)-function. Finally, in
section 7 we extend our study of the flows to the models
\(SM(2,4n+4)\) which are the generalisations of the SLY model.

\end{section}

\begin{section}{The supersymmetric Lee\,--Yang model}

The supersymmetric Lee\,--Yang model or the superconformal minimal model
SM(2,8) is a non-unitary conformal field theory with central charge
\(c=-\frac{21}4\). The super Kac-table with the highest weights of the
primary fields is

\begin{center}
\begin{tabular}{|c|c|c|c|c|c|c|}
\hline
0&\(-\frac{3}{32}\)&\(-\frac{1}4\)&\(-\frac{7}{32}\)&\(-\frac{1}4\)&\(-\frac{3}{32}\)&0\\
\hline
\end{tabular}
\end{center}

The algebra of superconformal transformations in the plane are
generated by two fields, \(T(z)\) and \(G(z)\) whose mode expansions
read
\begin{subequations}
\begin{align}
T(z)&=\sum_n L_n z^{-n-2}\,,\label{mode1}\\
G(z)&=\sum_r G_r z^{-r-3/2}\,.\label{mode2}
\end{align}
\label{modee}
\end{subequations}
The \(N=1\) supersymmetric extension of the Virasoro algebra is
defined by the following (anti)commutation relations:
\begin{subequations}
\begin{align}
[L_n,L_m]&=(n-m)L_{n+m}+\frac{c}{12}n(n^2-1)\delta_{n,-m}\,,\label{a1}\\
\{G_r,G_s\}&=2L_{r+s}+\frac{c}3\left(r^2-1/4\right)\delta_{r,-s}\,,\;\label{a2}\\
[L_n,G_r]&=\left(\frac{n}2-r\right)G_{n+r}\,.\label{a3}
\end{align}
\label{algebra}
\end{subequations}
Here \(n\), \(m\) denote integers, while depending on the boundary
conditions for \(G(z)\) the indices \(r\), \(s\) can take half-integer
(Neveu--Schwarz sector) or integer (Ramond sector) values.

The highest weight representations of the algebra are defined by 
highest weight states \(|\Delta\rangle\) satisfying
\begin{subequations}
\begin{align}
L_n|\Delta\rangle&=0\qquad\qquad n>0\,,\label{hw1}\\
L_0|\Delta\rangle&=\Delta\,,\label{hw2}\\
G_r|\Delta\rangle&=0\qquad\qquad r>0\,.\label{hw3}
\end{align}
\label{hw}
\end{subequations}
The Verma module is generated by operators having a nonpositive
index. The vectors
\begin{equation}
L_{n_1}\dots L_{n_k}G_{r_1}\dots G_{r_l}|\Delta\rangle\,,\qquad
n_1\le\dots\le n_k<0\,,\;\;r_1<\dots<r_l\le0
\label{basis}
\end{equation}
constitute a basis for the Verma module. Note that the inequalities
for the \(r_i\)-s are strict since
\(G_r^2=L_{2r}-\frac{c}{12}\,\delta_{r,0}\) by equation (\ref{a2}).

In this paper we consider the so-called spin model which can be
obtained by the projection onto the states with even fermion parity of
the Hilbert space.

One can consider this superminimal model on a strip of width \(R\)
with nontrivial boundary conditions at the edges of the strip. In the
Ramond sector the superconformal boundary conditions are in one-to-one
correspondence with the highest weight representations so they can be
indexed by the same label set (the Kac indices), whereas in the
Neveu--Schwarz sector there are two boundary conditions for each entry
of the Kac table, the NS and the \(\widetilde{\text{NS}}\) boundary
conditions \cite{nepomechie}. 

This model can also be defined as the minimal model M(3,8) with Kac
table

\begin{center}
\begin{tabular}{|c|c|c|c|c|c|c|}
\hline
0&-7/32&-1/4&-3/32&1/4&25/32&3/2\\
\hline
3/2&25/32&1/4&-3/32&-1/4&-7/32&0\\
\hline
\end{tabular}
\end{center}
Suitable combinations of the Virasoro representations constitute irreducible
representations of the superconformal algebra:
\begin{subequations}
\begin{align}
\bigx{1.3}(0\bigx{1.3})_\text{SVir}&\longleftrightarrow\bigx{1.3}(0\bigx{1.3})_\text{Vir}\oplus\bigx{1.3}(\frac32\bigx{1.3})_\text{Vir}\\
\bigx{1.3}(-\frac14\bigx{1.3})_\text{SVir}&\longleftrightarrow\bigx{1.3}(-\frac14\bigx{1.3})_\text{Vir}\oplus\bigx{1.3}(\frac14\bigx{1.3})_\text{Vir}\\
\bigx{1.3}(-\frac{7}{32}\bigx{1.3})_\text{SVir}&\longleftrightarrow\bigx{1.3}(-\frac{7}{32}\bigx{1.3})_\text{Vir}\oplus\bigx{1.3}(\frac{25}{32}\bigx{1.3})_\text{Vir}\,.\\
\bigx{1.3}(-\frac{3}{32}\bigx{1.3})_\text{SVir}&\longleftrightarrow\bigx{1.3}(-\frac{3}{32}\bigx{1.3})_\text{Vir}\,.
\end{align}
\end{subequations}
Since the superprimary fields are also Virasoro primary fields and the
\(g\)-factors are the same for the NS and \(\widetilde{\text{NS}}\)
boundaries, the \(g\)-factors of the various conformal boundary
conditions can be calculated in the Virasoro picture. The matrix
elements of the \(S\) modular transformation are
\begin{equation}
S_{(r,s),(r',s')}=2\sqrt{\frac2{pq}}(-1)^{rs'+r's+1}\sin
\left(rr'\frac{q}p\pi\right)\sin\left(ss'\frac{p}q\pi\right)
\end{equation}
and according to \cite{drtw} the \(g\)-factors are given by
\begin{equation}
g_{(1,s)}=\frac{S_{\Omega,(1,s)}}{\sqrt{|S_{\Omega,0}|}}
\end{equation}
where 0 denotes the conformal vacuum and \(\Omega\) denotes the state
of lowest conformal weight. In our case \(0=(1,1)\) and
\(\Omega=(1,3)\), so
\begin{equation}
g_{(1,s)}=\frac{(-1)^s}{\sqrt{2}}\frac{\sin\frac{9\pi}8s}{\sqrt{\sin\frac\pi8}}\,.
\end{equation}

If we choose the \(\beta=(1,1)\) boundary condition at one of the edges
and \(\alpha=(1,s)\) at the other one, the partition function on a
cylinder of circumference \(L\) and length \(R\) will be
\begin{multline}
\mathcal{Z}(R,L)=\mathrm{tr}_\text{NS}\frac{1+(-1)^F}2e^{-LH_{(1,1),(1,s)}}+
\mathrm{tr}_\text{R}\frac{1+(-1)^F}2e^{-LH_{(1,1),(1,s)}}=\\
\mathrm{tr}_{(1,s)}\frac{1\pm(-1)^F}2e^{-\pi\frac{L}{R}(L_0-c/24)}\,,
\label{hilbert}
\end{multline}
where the sign is ``\(+\)'' if both boundaries are of NS or
\(\widetilde{\text{NS}}\) type and ``\(-\)'' if they are different
(see \cite{nepomechie, kormos} for details). Thus the Hilbert space
consists of the {\em bosonic levels} of the corresponding
superconformal Verma module {\em if the boundary conditions are both
of NS or \(\widetilde{\text{NS}}\) type} and of the {\em fermionic
levels} of this module {\em if they are different}. For a bosonic
primary field the bosonic levels are the integer levels and for a
fermionic primary field they are the half-integer
levels.

We consider the boundary perturbation of this model defined by the
action:
\begin{equation}
\mathcal{A}=\mathcal{A}_\text{SM(2,8)}+h\int_{-\infty}^\infty\ud
  t\,(\hat G_{-1/2}\phi_{1,3})(0,t)\,.
\end{equation}
where \((\hat G_{-1/2}\phi)\) is the field appearing in the
operator product expansion of \(G(z)\) and \(\phi\):
\begin{equation}
G(z)\phi(w)=\frac{(\hat G_{-1/2}\phi)(w)}{z-w}+\dots\,
\end{equation}
This operator can live on the \((1,s)\) boundary since the fusion
coefficient \(N_{(1,3),(1,s)}^{(1,s)}\) is non-zero. This boundary
perturbation is relevant, since the conformal dimension of the
perturbing operator is \(1/4<1\). Since there is no perturbation in
the bulk, the theory remains massless along the flow. The perturbation
preserves supersymmetry so the boundary renormalisation group flows
will take place in the space of possible supersymmetric boundary
conditions, in which the fixed points are the superconformal
ones. This perturbation is also integrable \cite{mathieu}.

\end{section}

\begin{section}{The reflection factor}
We obtain the S-matrix and the reflection factor of the supersymmetric
Lee\,--Yang model by considering it as a reduction of the
supersymmetric sine--Gordon (SSG) model. The original idea is of
Smirnov \cite{smirnov} and it was first applied in supersymmetric
context for bulk S-matrices in \cite{ahn90}. The SSG S-matrix was also
derived in \cite{ahn90}, the formulae are collected in appendix
\ref{ssgs}.

If \(\lambda>2\) in the sine--Gordon model the pole in the S-matrix
describing the scattering of the first breather on itself corresponds
to the second breather as a bound state. For \(\lambda<2\) only the
first breather is in the spectrum and the pole is explained by the
Coleman--Thun mechanism \cite{colemanthun} which requires the presence
of solitons in the theory. However, at the particular value of the
coupling constant \(\lambda=\frac32\) the masses and the S-matrices
for the first and the would-be second breathers become equal:
\(m_1=m_2\) and
\(S^{1,1}_\text{SSG}=S^{1,2}_\text{SSG}=S^{2,2}_\text{SSG}\), so the
pole can be explained by the self fusion of the only breather and the
solitons can be projected out consistently from the theory
\cite{ahn90}. Thus the SSG model is reduced to the supersymmetric
scaling Lee\,--Yang model containing only one superdoublet. The
S-matrix simplifies to
\begin{equation}
S_\text{SLY}(\theta)=S_\text{LY}(\theta)S_\text{SUSY}(\theta)
\end{equation}
where
\begin{equation}
S_\text{LY}(\theta)=\frac{\sinh(\theta)+i\sin(\frac{2\pi}3)}{\sinh(\theta)-i\sin(\frac{2\pi}3)}
\end{equation}
and
\begin{gather}
S_\text{SUSY}(\theta)=\tilde{Z}(\theta)\times
\begin{pmatrix}
1+2i\,\frac{\sin(\frac\pi3)}{\sinh(\theta)}&0&0&\frac{\sin(\frac\pi3)}{\cosh(\frac\theta2)}\\
0&1&i\,\frac{\sin(\frac\pi3)}{\sinh(\frac\theta2)}&0\\
0&i\,\frac{\sin(\frac\pi3)}{\sinh(\frac\theta2)}&1&0\\
\frac{\sin(\frac\pi3)}{\cosh(\frac\theta2)}&0&0&-1+2i\,\frac{\sin(\frac\pi3)}{\sinh(\theta)}
\end{pmatrix}\,,
\label{ssusy}\\
\tilde{Z}(\theta)=\frac{\sinh(\frac\theta2)}{\sinh(\frac\theta2)+i\sin(\frac\pi3)}\,\exp\left\{\int_0^\infty\frac{\ud
  t}t\frac{\sinh(\frac{t}3)\sinh(\frac{2t}3)}{\cosh^2(\frac{t}2)\cosh(t)}\sinh(\frac{i
  t\theta}{\pi})\right\}\,.
\end{gather}
This is the S-matrix used also by Ahn and Nepomechie in \cite{an}. For
later convenience we introduce the notation
\begin{equation}
Z(\theta)=S_\text{LY}(\theta)\tilde{Z}(\theta)\,.
\label{zdef}
\end{equation}

It is easy to check that this S-matrix satisfies the appropriate
unitarity and crossing relations and that the Yang--Baxter equations
are also satisfied.

\begin{subsection}{The ``folding''}
We have seen that at the particular coupling \(\lambda=\frac32\) the
bulk SSG model reduces to the SLY model. For this reduction to be
consistent also in the boundary case it is necessary that at this
point the reflection factors for the 1st and the would-be 2nd breather
be equal. This requirement gives functional relations between the two
SSG boundary parameters, \(\eta\) and \(\vartheta\) (first considered
in \cite{bssgrevisited}).

The reflection factors of the boundary SSG model can be found in
appendix \ref{ssgr}. In the \(BSSG^+\) case the supersymmetric factor
\(R_\text{SUSY}(\theta)\) (see \ref{rsusy}) does not contain the
boundary parameters and the only way it depends on the species of the
breather is through the parameter \(\rho\) which at this special
coupling is \(\rho_k=\pi-k\frac{2\pi}3\). From
\(\rho_1=-\rho_2=\frac\pi3\) it is obvious that
\(\mathcal{A}_\pm(\theta)\) are automatically the same for \(k=1,2\).

Now we turn to the scalar part, i.\ e.\ to the SG reflection
factors. It turns out that \(R^{(1)}(\theta)=R^{(2)}(\theta)\) if
\begin{equation}
\eta=i\vartheta+\frac{2k+1}2\pi\,,\qquad\qquad k\in\mathbb{Z}\,.
\end{equation}
If \(k=1\), i.\ e. \(\eta=i\vartheta+\frac{3\pi}2\) then
\(R^{(1)}=R^{(2)}=R_0\) where
\begin{equation}
R_0(\theta)=\left(\frac12\right)\left(\frac32\right)\left(\frac42\right)^{-1}\,,
\end{equation}
where we introduced the notation
\[(x)=\frac{\sin(\frac{\theta}{2i}+x\frac\pi6)}{\sin(\frac{\theta}{2i}-x\frac\pi6)}\,.\]
If \(k=1\), i.\ e.\ \(\eta=i\vartheta+\frac\pi2\) then \(R^{(1)}=R^{(2)}=R_I\) with
\begin{multline}
R_I(\theta)=R_0(\theta)\left(\frac{1+b}2\right)\left(\frac{1-b}2\right)^{-1}\left(\frac{5-b}2\right)\left(\frac{5+b}2\right)^{-1}=\\
R_0(\theta)\left(S_\text{LY}\bigx{1.2}(\theta+i\frac{b+3}6\pi\bigx{1.2})S_\text{LY}\bigx{1.2}(\theta-i\frac{b+3}6\pi\bigx{1.2})\right)^{-1}\,,
\end{multline}
where we introduced the parameter
\(b=\frac{4i}\pi\vartheta-2\). \(R_0(\theta)\) and \(R_I(\theta)\) are
the same reflection factors as those given by Dorey et al. (for
example in \cite{bcolemanthun}).

Finally, if \(k=-1\), i,\ e.\ \(\eta=i\vartheta-\frac\pi2\) then
\(R^{(1)}=R^{(2)}=R_{II}\) where
\begin{multline}
R_{II}(\theta)=R_0(\theta)\left(\frac{b-1}2\right)\left(\frac{b+3}2\right)^{-1}\left(\frac{b-3}2\right)\left(\frac{5+b}2\right)^{-1}=\\
R_0(\theta)\left(S_\text{LY}\bigx{1.2}(\theta+i\frac{b+1}6\pi\bigx{1.2})S_\text{LY}\bigx{1.2}(\theta-i\frac{b+1}6\pi\bigx{1.2})\right)^{-1}\,.
\end{multline}
The other values of \(k\) do not lead to new reflection factors. With
the redefinition \(b\to b+4\) the factor \(R_{II}(\theta)\) becomes
\begin{multline}
R_{II}(\theta)=R_0(\theta)\left(\frac{b+3}2\right)\left(\frac{5-b}2\right)\left(\frac{b+1}2\right)\left(\frac{3-b}2\right)=\\
R_0(\theta)\left(S_\text{LY}\bigx{1.2}(\theta+i\frac{b+5}6\pi\bigx{1.2})S_\text{LY}\bigx{1.2}(\theta-i\frac{b+5}6\pi\bigx{1.2})\right)^{-1}\,,
\end{multline}
which can be identified as the reflection factor for the excited
boundary \cite{bcolemanthun}. Thus \(R_0\) and \(R_I\) are the two
ground state reflection factors for the two boundary conditions of the
Lee\,--Yang model. They satisfy the boundary unitarity and boundary
crossing equations:
\begin{align}
R(\theta)R(-\theta)&=1\,,\\
R\bigx{1.3}(\frac{i\pi}2-\theta\bigx{1.3})&=S_\text{LY}(2\theta)R\bigx{1.3}(\frac{i\pi}2+\theta\bigx{1.3})\,.
\end{align}
The SLY reflection factors also satisfy the boundary unitarity
equation provided \(\mathcal{A}_\pm(\theta)\mathcal{A}_\pm(-\theta)=1\) and this is indeed
true:
\begin{equation}
\mathcal{A}_\pm(\theta)\mathcal{A}_\pm(-\theta)=2\frac{\cos(\frac{\theta}{2i}\mp\frac\pi4)\cos(-\frac{\theta}{2i}\mp\frac\pi4)}{\cosh(\theta)}
=1\,.
\end{equation}
The boundary crossing equations are
\begin{subequations}
\begin{gather}
2^{-\frac{2\theta}{i\pi}}R_\phi^\phi\bigx{1.3}(\frac{i\pi}2-\theta\bigx{1.3})=S_{\phi\phi}^{\phi\phi}(2\theta)R_\phi^\phi\bigx{1.3}(\frac{i\pi}2+\theta\bigx{1.3})
+S_{\psi\psi}^{\phi\phi}(2\theta)R_\psi^\psi\bigx{1.3}(\frac{i\pi}2+\theta\bigx{1.3})\,,\\
2^{-\frac{2\theta}{i\pi}}R_\psi^\psi\bigx{1.3}(\frac{i\pi}2-\theta\bigx{1.3})=S_{\psi\psi}^{\psi\psi}(2\theta)R_\psi^\psi\bigx{1.3}(\frac{i\pi}2+\theta\bigx{1.3})
+S_{\phi\phi}^{\psi\psi}(2\theta)R_\phi^\phi\bigx{1.3}(\frac{i\pi}2+\theta\bigx{1.3})\,.
\end{gather}
\label{bcross}
\end{subequations}
Since the Lee\,--Yang parts satisfy the scalar version of the boundary
crossing condition we only have to check that the supersymmetric
factors \(S_\text{SUSY}\) and \(R_\text{SUSY}\) satisfy
\eqref{bcross}. Using the integral representations for them this can
be proved analytically. We also checked that the matrix parts satisfy
the boundary Yang--Baxter equations as well.

It seems plausible that the two supersymmetric reflection factors,
\(R_0(\theta)R_\text{SUSY}(\theta)\) and
\(R_I(\theta)R_\text{SUSY}(\theta)\) are the reflection factors for
the two Neveu--Schwarz boundary conditions of the supersymmetric
scaling Lee\,--Yang model, since for the Ramond type boundary
conditions we do not expect diagonal reflection factors (the fermion
parity can change). It is also natural to think that
\(R_0R_\text{SUSY}\) is for the (1,1) boundary condition, because it
does not depend on any boundary parameter.

We emphasize that the reflection factors proposed here are different
from the non-diagonal one proposed by Ahn and Nepomechie in \cite{an}.

\end{subsection}

\begin{subsection}{Relation between the boundary parameter \(b\) and the\\
    boundary coupling \(h\)} 

Following reference \cite{dptw} it is possible to conjecture the
relation between \(b\) and the coupling constant of the boundary
perturbation, \(h\) (a similar argument was used also in
\cite{bpttgzs}). In that paper it was found that the transformation
\(b\to4-b\) is a symmetry of the model, since this change of \(b\)
maps the reflection factors of the ground state and excited boundaries
into each other. For \(b>2\) the pole at \(\theta=\frac{i(b+1)\pi}6\)
in \(R_I\) which corresponds to an excited boundary leaves the
physical strip but at the same time the pole at
\(\theta=\frac{i(5-b)\pi}6\) in \(R_{II}\) enters. Thus for \(b>2\)
\(R_{II}\) describes the ground state boundary and \(R_I\) the excited
boundary. A similar phenomenon is probably to occur in our case, but
exactly the same thing can not work, since now the excited boundary
state is a doublet while the boundary ground state is a singlet and
\(R_{II}\) is a 4x4 matrix. However, at the pole \(R_{II}\) becomes a
1-dimensional projector. Since the reflection factors are manifestly
periodic in \(b\) with a period of 12 there is a point (\(b=8\))
beyond which the boundary energies swap back and the ground state will
be a singlet again. This gives a symmetry \(b\to10-b\). Together with
the trivial symmetry \(b\to-6-b\) of the reflection factors this
results in a periodicity of the physical spectrum in \(b\) with a
period of 16. Similarly to the case in \cite{dptw}, the value at half
way between the fixed points of the two symmetries should correspond
to \(h=0\). This leads to the conjecture for the relation between
\(b\) and \(h\):
\begin{equation}
h=h_c\sin\bigx{1.3}(\frac{(b-1)\pi}8\bigx{1.3})
\label{bhrel}
\end{equation}
with \(h_c=\alpha m^{3/4}\) where \(\alpha\) is some numeric constant
and \(m\) is some mass scale of the model, e.\ g.\ the mass of the
particle.
\end{subsection}

\begin{subsection}{The massless limit}
It is the supersymmetric part \(R_\text{SUSY}\) for which it is easier
to calculate the massless limit since it does not contain boundary
parameters. All we have to do is to take the limit
\(\theta\to\infty\).

\(\mathcal{A}_\pm(\theta)\) can be written in the following form:
\begin{equation}
\mathcal{A}_\pm(\theta)=\sqrt{2}2^{-\frac\theta{i\pi}}\frac{\cos(\frac\theta{2i}\mp\frac\pi4)}{\sqrt{\cosh(\theta)}}\,e^{\,I(\theta)}\,,
\end{equation}
where 
\begin{equation}
I(\theta)=\int_0^\infty\frac{\ud t}t f(t)\sinh(\frac\theta{i\pi}t)=
-\frac12\int_{-\infty}^\infty\frac{\ud t}t e^{i t\theta}f(t\pi)
\end{equation}
with
\begin{equation}
f(t)=-\frac14\frac{\cosh(\frac{t}6)+\cosh^2(\frac{t}2)}{\cosh^2(\frac{t}4)\cosh^2(\frac{t}2)}\,.
\end{equation}
Now
\begin{equation}
I(\infty)=-\frac{i\pi}2f(0)=\frac14i\pi
\end{equation}
so
\begin{equation}
\lim_{\theta\to\infty}2^{\frac\theta{i\pi}}\mathcal{A}_\pm(\theta)=\sqrt{2}\,\frac{\frac12\, 
      e^{\frac\theta2}\, e^{\mp i\frac\pi4}}{\sqrt{\frac12\,
      e^\theta}}\, e^{\frac{i\pi}4}=e^{\frac{i\pi}4(1\mp1)}\,,
\end{equation}
which means that
\begin{equation}
R_\text{SUSY}(\theta)\to\begin{pmatrix}1&0\\0&i\end{pmatrix}\,.
\end{equation}

Let us turn to the scalar parts. \(R_0(\theta)\) does not depend on
the boundary parameters and
\begin{equation}
\lim_{\theta\to\infty} R_0(\theta)=1\,.
\end{equation}
\(R_I(\theta)\) does depend on the boundary parameter \(b\). In the
massless limit we want to keep \(h\) fixed, which means that the sine
in \eqref{bhrel} has to go to infinity in a proper way. This can be
achieved only if \(b\) becomes complex. We should set \(b=-3-i\hat
b\), then
\begin{equation}
h=-h_c\cosh(\frac{\hat{b}\pi}8)m^{\frac34}\longrightarrow
-\frac{h_c}2(e^{\frac{\hat{b}\pi}6}m)^{\frac34}\,.
\end{equation}
This means that \(m\, e^{\frac{\hat{b}\pi}6}\) should be kept fixed
while \(m\to0\) and \(\hat{b}\to\infty\). We also want to keep the
physical momentum, \(m\cosh(\theta)\) fixed, which implies that
\(me^\theta\) is constant.  We conclude that while \(\theta,\,
b\to\infty\) the combination \(\theta-\frac{\hat b}6\pi\) is fixed.
So
\begin{equation}
R_I(\theta)=R_0(\theta)\left(S_\text{LY}\bigx{1.2}(\theta+\frac{\hat{b}}6\pi\bigx{1.2})S_\text{LY}\bigx{1.2}(\theta-\frac{\hat{b}}6\pi\bigx{1.2})\right)^{-1}\longrightarrow
S_\text{LY}^{-1}\bigx{1.2}(\theta-\frac{\hat{b}}6\pi\bigx{1.2})=S_\text{LY}^{-1}(\hat\theta-\theta_\text{B})
\end{equation}
where \(\hat\theta\) is the pseudo-rapidity and \(\theta_\text{B}\) is
related to \(\hat{b}\).\,\footnote{The method used in \cite{an}, taking
  the joint limits \(\theta\to\infty\), \(m\to0\) and demanding that
  the result be finite and unitary, leads to the same result.}

We have found that the massless reflection factors for the two NS
boundary conditions of the SLY model are
\begin{equation}
\hat{R}_0=\begin{pmatrix}1&0\\0&i\end{pmatrix}\quad\text{and}\quad
  \hat{R}_I(\hat\theta)=S_\text{LY}^{-1}(\hat\theta-\theta_\text{B})\hat{R}_0\,.
\label{Rs}
\end{equation}
In the following we will omit the hat in the notation for the pseudo-rapidity.
\end{subsection}

\end{section}

\begin{section}{TCSA}
The method we use for examining boundary flows is the so-called
truncated conformal space approach, or TCSA. In this approach the
infinite dimensional Hilbert space is truncated to a finite
dimensional vector space by using only those states whose energy is
not greater than a threshold value, \(E_\text{cut}\). This is
equivalent to truncating the Hilbert space at a given level. The
Hamiltonian is then diagonalised on this truncated space. The original
idea was proposed in \cite{yurovzam}, it was applied for the first
time for boundary problems in \cite{dptw} and in a supersymmetric
theory in \cite{susytcsa}. The current method is based on the
techniques of \cite{kausch}, modified for superconformal minimal
models for the first time in \cite{kormos}.

The Hamiltonian of the perturbed superconformal field theory takes the
form
\begin{equation}
\hat{H}=\hat{H}_\text{CFT}+\hat{H}_\text{pert}\,,
\end{equation}
where \(\hat{H}_\text{CFT}\) is the Hamiltonian of the unperturbed theory
\begin{equation}
\hat{H}_\text{CFT}=\hat{H}_{\alpha0}=\frac\pi{R}(L_0-\frac{c}{24})
\end{equation}
and \(\hat{H}_\text{pert}\) comes from the perturbation on the strip:
\begin{equation}
\hat{H}_\text{pert}^\text{strip}=h\,(\hat G_{-1/2}\phi_{1,3})(0,0)\,.
\end{equation}
The location of the left boundary is at \(x=0\) and we are free to
choose \(t=0\) for calculating the spectrum. By the exponential map
this on the \(z\)-plane becomes
\begin{equation}
\hat{H}_\text{pert}=h\,\left(\frac\pi{R}\right)^{\Delta_{1,3}+1/2}\bigx{1.2}(\hat G_{-1/2}\phi_{1,3}\bigx{1.2})(z=1)\,.
\end{equation}
Thus the complete Hamiltonian on the plane can be written as
\begin{equation}
\hat{H}=\frac\pi{R}\left[L_0-\frac{c}{24}+h\left(\frac{R}\pi\right)^{1/2-\Delta_{1,3}}(\hat{G}_{-1/2}\phi_{1,3})(1)\right]\,.
\end{equation}
Since a numerical calculation requires dimensionless quantities we
have to introduce some mass scale, \(\mu\) and measure the volume
(\(R\)) and the energies in units of \(\mu\). In other words, we use
the dimensionless quantities \(r=\mu R\), \(\varepsilon=E/\mu\),
\(\kappa=h/\mu^{1/2-\Delta_{1,3}}\) and \(\hat{h}=\hat{H}/\mu\):
\begin{equation}
\hat{h}=\frac\pi{r}\left[L_0-\frac{c}{24}+\kappa\left(\frac{r}\pi\right)^{1/2-\Delta_{1,3}}(\hat{G}_{-1/2}\phi_{1,3})(1)\right]\,,
\label{hamilton}
\end{equation}
where \(\kappa\) characterizes the relation between the boundary
coupling and the chosen mass scale, thus it is not fixed to any
preferred value yet.

In order to calculate the eigenvalues of the matrix \(\hat{h}\) we
choose the basis vectors for the truncated Hilbert space of the form
written in \eqref{basis} with \(\Delta=\Delta_{1,s}\). After
eliminating the null vectors and the corresponding submodules, we
obtain a non-orthonormal basis, \(\{e_i\}\). Then the matrix elements
of the TCSA Hamiltonian are
\begin{equation}
h_{ij}=\frac\pi{r}\left[(h_i-\frac{c}{24})\delta_{ij}+\kappa\left(\frac{r}\pi\right)^{1/2-\Delta_{1,3}}(M^{-1}B)_{ij}\right]\,,
\end{equation}
where \(M\) is the inner product matrix
\begin{equation}
M_{ij}=\langle e_i|e_j\rangle\,,
\end{equation}
and \(B\) contains the matrix elements of the perturbing operator:
\begin{equation}
B_{ij}=\langle e_i|(\hat G_{-1/2}\phi_{1,3})(1)|e_j\rangle\,.
\end{equation}
These matrix elements can be calculated using contour integration
techniques and the superconformal operator algebra, as worked out in
detail in \cite{kormos}.

\begin{subsection}{TCSA fits}
The Bethe--Yang equation for a massless particle moving between the
boundaries is
\begin{equation}
e^{i p 2R}\hat{R}_0\hat{R}_I(\theta)=1\,,
\end{equation}
that is
\begin{equation}
e^{i p 2R}S_\text{LY}^{-1}(\theta-\theta_\text{B})\hat{R}_0^2(\theta)=1\,.
\end{equation}
Taking the logarithm we get
\begin{equation}
2pR+\delta(\theta-\theta_B)=2n\pi\,,
\label{by}
\end{equation}
where
\begin{equation}
\delta(\theta)=\frac{1}i\log S_\text{LY}^{-1}(\theta)\,,
\label{delta}
\end{equation}
and \(n\) can be integer or half-integer for the two eigenvalues of
\(\hat{R}_0^2\) (1 and \(-1\)).

For a massless particle \(p=E=\frac\mu2e^\theta\) where \(\mu\) is
some mass scale. A shift in the pseudo-rapidity \(\theta\) is
equivalent with a change in \(\mu\). The shift in \(\theta\),
\(\theta-\theta_\text{B}\) can be compensated by changing \(\mu\)
which for the TCSA means changing the numeric value of the
dimensionless coupling \(\kappa\).

We should note that the Bethe--Yang equations are not derived
rigorously for massless particles. However, experience shows that it
works and gives the correct energy levels even for rather small
volumes. The massless limit of the massive TBA equations are the same
as those derived from the massless Bethe--Yang equation and they work
in a lot of cases, which is another reassuring fact. We will see that
also in our case it gives consistent energy levels.

Equation \eqref{by} can be solved for \(r\) (\(pR=ER=\varepsilon r\))
for different values of \(n\) and we can plot the inverse of the
function \(r(\varepsilon)\) and compare the lines with the TCSA data,
\(\varepsilon_\text{TCSA}(r)\). For this we have to distinguish the
one-particle energy levels. For the multi-particle states we have a
coupled system of Bethe--Yang equations that contain also the bulk
S-matrix, which makes the behaviour of the energy eigenvalues
different for finite volumes. This, in principle, allows us to select
the one-particle levels. Furthermore, for large volume (\(r\)) the
interaction becomes negligible and the asymptotic behaviour of the
energies are
\begin{equation}
E_n\sim\frac\pi{r}n
\end{equation}
for one-particle states and
\begin{equation}
E_{n_1,n_2,\dots}\sim\frac\pi{r}\sum_i n_i
\end{equation}
for multi-particle states. Thus the IR behaviour is the same for
one-particle and multi-particle states and for a given (half-)integer
number \(n\) there is precisely one one-particle state and possibly
several multi-particle states with asymptotic energy \(n\pi/r\). If
for a certain value of \(n\) there is only one level, one can be sure
that it is a one-particle level. For \(n=\frac12\) there are no
multi-particle states obviously and for greater values of \(n\)
certain multi-particle states may be forbidden by exclusion rules. For
example in the \(n=1\) case the two-particle state corresponding to
\(1=\frac12+\frac12\) is excluded, so for \(n=1/2,\,1\) there is only
one energy level which corresponds to a one-particle state.

For each value of \(\kappa\) first we must find the appropriate value
of \(\theta_\text{B}\), for example by fitting the lowest energy level
which is a one particle level. After this is done, all the other one
particle energy eigenvalues should automatically fit the other
lines. The result of such a fit can be seen in Figure \ref{tcsaint}
for the NS-NS case with \(\kappa>0\), when only the integer levels of
the module (1,3) are in the Hilbert space (see \eqref{hilbert} and
below). As can be seen, the TCSA spectrum possesses the expected
features: the lowest energy levels are non-degenerate in the IR and
the higher levels are arranged in groups.

In every group there is exactly one level that fits the one-particle
Bethe--Yang energy, with integer values \(n\). It is interesting to
observe that these levels are always the highest ones in their group,
which is due to the fact that the interaction is attractive. The lines
of Figure \ref{tcsaint} not fitted by a solid line correspond to
multi-particle states. The number of these levels is neither
consistent with bosonic nor with fermionic exclusion statistics. This
suggests that these particles obey some generalised exclusion
statistics, similarly to the ordinary Lee\,--Yang model \cite{excl}.

One would expect that the other eigenvalue of the reflection factor
corresponds to the reflection of the other member of the superdoublet,
that is the energy eigenvalues from the half-integer levels
(NS-\(\widetilde{\text{NS}}\) case) will fit the lines of half-integer
values of \(n\). Surprisingly, this is not the case and it is not
clear why it does not happen and then what the reflection factor is
for these fermionic levels. 

For \(\kappa<0\) (\(h<0\)) the spectrum becomes complex, just like for
the ordinary Lee\,--Yang model \cite{dptw}.

Another and more precise approach is obtaining the phase shift
\(\delta(\theta)\) from the various TCSA eigenvalues and comparing it
with the exact function \eqref{delta}. From \eqref{by} one gets
\begin{equation}
\delta_\text{TCSA}(\log\frac{E_\text{TCSA}}\mu-\theta_\text{B})=
\delta_\text{TCSA}(\log\varepsilon_\text{TCSA}-\theta_\text{B})=
2n\pi-2\varepsilon\,r\,, 
\end{equation}
It can be seen that our ignorance of the proper mass scale (or the
value of \(\theta_\text{B}\)) is only a matter of a horizontal shift
of the function along the \(\theta\)-axis while changing the quantum
number \(n\) shifts the function vertically. In Figure \ref{allint}
the phase shifts calculated from the lowest one particle eigenvalues
are plotted onto each other, which shows that the phase shifts
extracted from the different one-particle levels are the same. This
shows that the massless Bethe--Yang equation is meaningful and gives
consistent results. The difference between the phase shifts for small
rapidities is due to the truncation errors of the TCSA in the IR.

In Table \ref{deltatabl} the phase shift values calculated from the
difference of the two lowest TCSA eigenvalues are compared with the
theoretical values \eqref{delta}. In Figures \ref{dfit2} and
\ref{dfit3} the the TCSA phase shifts obtained from the lowest
eigenvalues are plotted together with the proposed phase shift
\eqref{delta}. As it can be seen there is good agreement.


\begin{table}[h!]
\begin{center}
\begin{tabular}{|c|c|c|}
\hline
\(\theta_\text{TCSA}\)&\(\delta(\theta)\)&\(\delta_\text{TCSA}(\theta)\)\\
\hline
\hline
0.0025  & 3.1358    & 3.0401 \\
\hline
0.0570   & 3.0099   & 2.9276 \\
\hline
0.1154   & 2.8760    & 2.8076 \\
\hline
0.1782   & 2.7337    & 2.6796 \\
\hline
0.2462   & 2.5824    & 2.5426 \\
\hline
0.3204    & 2.4217   & 2.3961 \\
\hline
0.4022   & 2.2513   & 2.2391 \\
\hline
0.4935    & 2.0707   & 2.0707 \\
\hline
0.5968    & 1.8795   & 1.8898 \\
\hline
0.7158    & 1.6772    & 1.6953 \\
\hline
0.8562    & 1.4630   & 1.4860 \\
\hline
1.0275    & 1.2361   & 1.2606 \\
\hline
1.2461    & 0.9951   & 1.0179 \\
\hline
1.5461    & 0.7378    & 0.7567 \\
\hline
2.0163    & 0.4612   & 0.4765 \\
\hline
3.0696    & 0.1609  & 0.1765 \\
\hline
\end{tabular}
\end{center}
\caption{The theoretical phase shift \eqref{delta} and the TCSA result}
\label{deltatabl}
\end{table}

\begin{figure}[h!]
\centering
\subfigure[Energy eigenvalues from the integer levels with the
  theoretical energies from the reflection factor]{\resizebox{72mm}{!}{\includegraphics{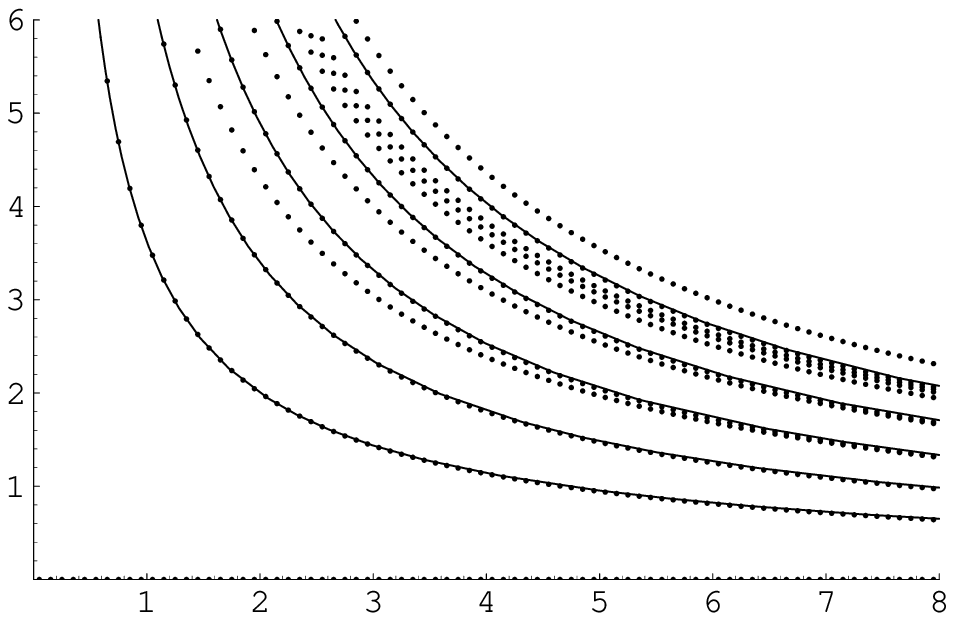}}\label{tcsaint}}
\hspace{2mm}\subfigure[\(\delta_\text{TCSA}(\theta)\) phase shifts
calculated from different TCSA
eigenvalues]{\resizebox{72mm}{!}{\includegraphics{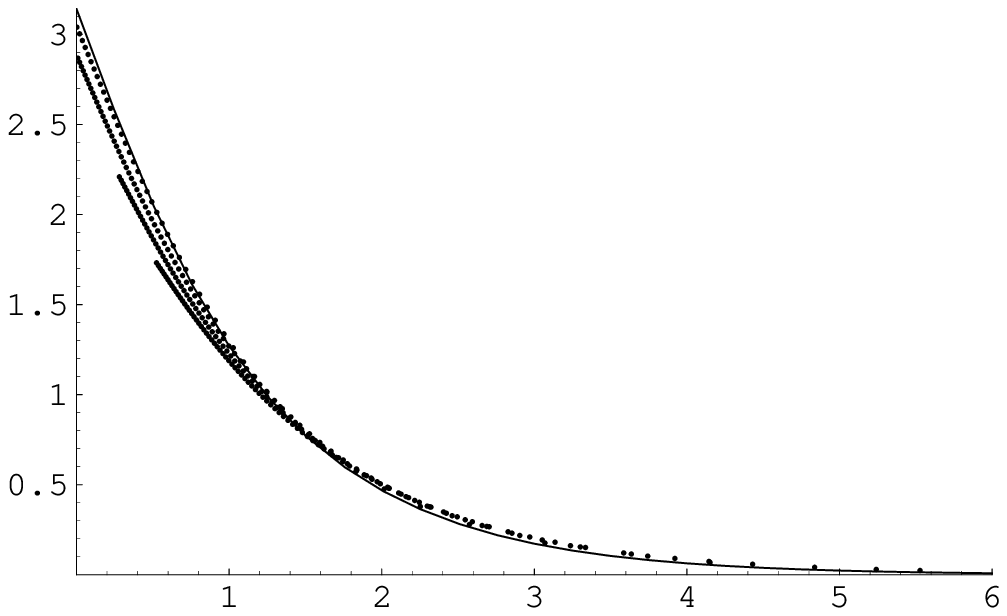}}\label{allint}}
\end{figure}

\begin{figure}[h!]
\centering \subfigure[TCSA phase shift from the difference of the
first two eigenvalues]{\resizebox{72mm}{!}{\includegraphics{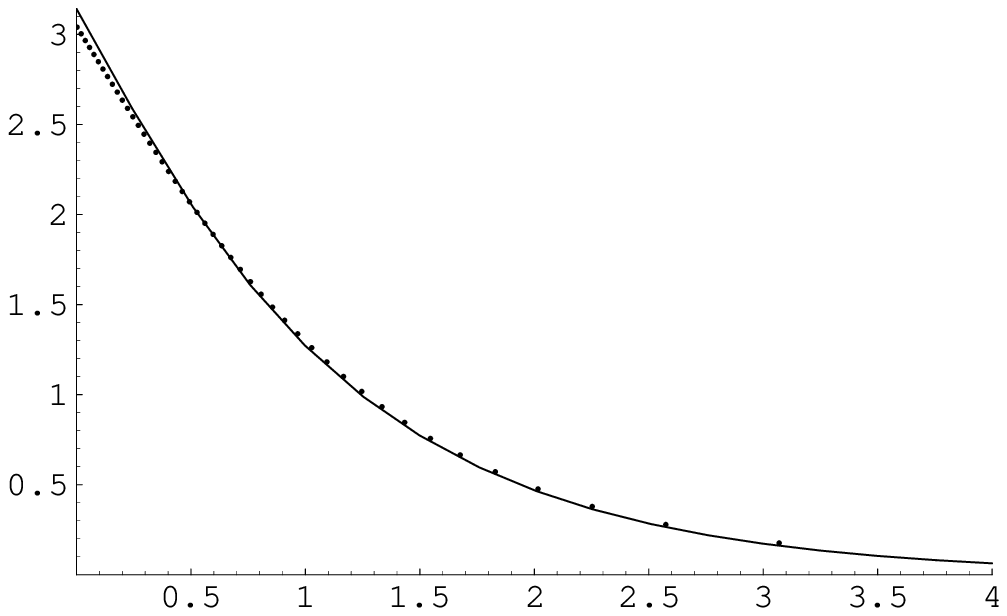}}\label{dfit2}}
\hspace{2mm}\subfigure[TCSA phase shift from the difference of the 3rd
and 1st eigenvalues]{\resizebox{72mm}{!}{\includegraphics{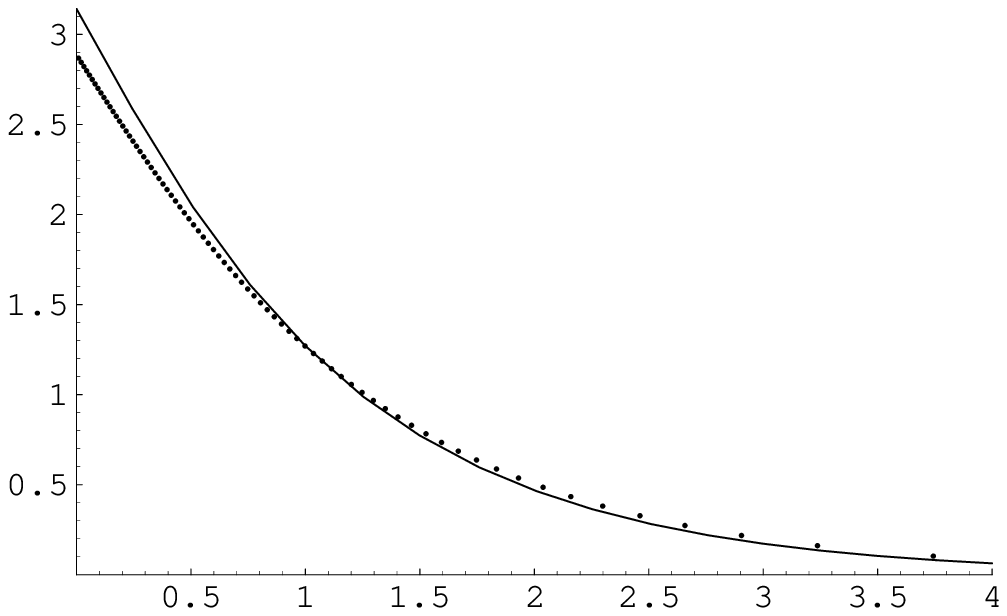}}\label{dfit3}}
\caption{Reflection factor and TCSA spectrum}
\end{figure}

\end{subsection}

\end{section}


\begin{section}{TBA}
The Thermodynamical Bethe Ansatz is based on the quantisation of the
energy levels using the S-matrices and the reflection factors of the
model. The corresponding Bethe--Yang equation takes the following form
in terms of the densities for \(N\) particles in the periodic case
(\cite{ahntba,moriconi}):
\begin{equation}
2\pi
P(\theta)=m\cosh(\theta)+\int\ud\theta'\bigx{1.8}[\rho(\theta')\Phi_z(\theta-\theta')\\
+\rho_0(\theta')\frac12\Phi(\theta-\theta')+\bar\rho_0(\theta')(-\frac12)\Phi(\theta-\theta')\bigx{1.8}]\,,
\end{equation}
where \(\rho(\theta)\) denote the density of the occupied states and
\(P(\theta)\) the density of available states. Similarly,
\(\rho_0(\theta)\) denotes the density distribution of real solutions
\(x_j\) of the equation
\begin{equation}
\prod_{l=1}^N\frac{\tanh(\frac{x-\theta_l}2-\frac{i\pi}6)}{\tanh(\frac{x-\theta_l}2+\frac{i\pi}6)}=-F\,,
\label{xeq}
\end{equation}
(\(F\) is the fermion parity) for which \(\epsilon_j=+1\) and by
\(\bar\rho_0(\theta)\) the solutions for which \(\epsilon_j=-1\). 

The kernel functions are
\begin{subequations}
\begin{align}
\Phi_z(\theta)&=\df{}{\theta}\text{Im}\log\left(\frac{Z(\theta)}{\sinh(\theta)}\right)\,,\\
\Phi(\theta)&=\df{}{\theta}\text{Im}\log S_\text{LY}(\theta)=-\frac{4\sqrt{3}\cosh(\theta)}{1+2\cosh(2\theta)}\,,
\end{align}
\label{phidef}
\end{subequations}
where \(Z(\theta)\) is the scalar part of \(S_\text{SLY}(\theta)\)
defined in equation \eqref{zdef}.

Let us turn to the boundary case. Since the reflection factors are
diagonal they do not affect the calculation of the transfer matrix
eigenvalues, so in the massless case
\begin{multline}
2\pi
P(\theta)=\frac\mu2 e^{\theta}+\frac{\Psi(\theta)}{2R}+\int\ud\theta'\bigx{1.8}[\rho(\theta')\Phi_z(\theta-\theta')\\
+\rho_0(\theta')\frac12\Phi(\theta-\theta')+\bar\rho_0(\theta')(-\frac12)\Phi(\theta-\theta')\bigx{1.8}]\,,
\label{peq}
\end{multline}
where
\begin{equation}
\Psi(\theta)=\df{}{\theta}\text{Im}\log(\pm \hat
R_\text{left}(\theta-\theta_\text{B}) \hat R_\text{right}(\theta-\theta_\text{B}))\,.
\label{psidef}
\end{equation}
For \(P_0(\theta)=\rho_0(\theta)+\bar\rho_0(\theta)\) we have a
similar equation, obtained from equation \eqref{xeq}:
\begin{multline}
2\pi P_0(\theta)=\int\ud\theta'\rho(\theta)\df{}{\theta}\text{Im}\log
\left(\frac{\tanh(\frac12(\theta-\theta'-i\frac\pi3))}{\tanh(\frac12(\theta-\theta'+i\frac\pi3))}\right)=\\
-\int\ud\theta'\rho(\theta)\Phi(\theta-\theta')\,.
\label{p0eq}
\end{multline}
Now using the fact that
\(P_0(\theta)=\rho_0(\theta)+\bar\rho_0(\theta)\) and equation
\eqref{p0eq} we can eliminate \(\rho_0(\theta)\) from equation
\eqref{peq} getting
\begin{equation}
P(\theta)=\frac{\mu}{4\pi}e^{\theta}+\frac{\Psi(\theta)}{4\pi R}+
\bigx{1.2}(\rho\star(\Phi_z-\frac12\Phi\star\Phi)\bigx{1.2})(\theta)
-(\bar\rho_0\star\Phi)(\theta)\,,
\end{equation}
where we introduced the notation
\begin{equation}
(\phi\star\psi)(\theta)=\int\frac{\ud \theta'}{2\pi}\phi(\theta-\theta')\psi(\theta')
\end{equation}
for the convolution. It turns out that the following identity holds (\cite{an}):
\begin{equation}
(\Phi_z-\frac12\Phi\star\Phi)(\theta)=\Phi(\theta)\,.
\end{equation}

The free energy of the system is given by
\begin{multline}
f=\int\ud\theta\bigx{1.3}\{\rho(\theta)\mu e^{\theta}\\
-T\bigl[P(\theta)\log P(\theta)-\rho(\theta)\log\rho(\theta)
  -(P(\theta)-\rho(\theta))\log(P(\theta)-\rho(\theta))\bigr]\\
-T\big[P_0(\theta)\log P_0(\theta)-\bar{\rho}_0(\theta)\log\bar{\rho}_0(\theta)
  -(P_0(\theta)-\bar{\rho}_0(\theta))\log(P_0(\theta)-\bar{\rho}_0(\theta))\bigr]\bigx{1.3}\}\,.
\end{multline}
Now the densities \(\rho(\theta)\), \(\bar{\rho}_0(\theta)\) can be varied to
minimize the free energy. Using
\begin{align}
\delta P&=\delta\rho\star\Phi-\delta\bar{\rho}_0\star\Phi\,,\\
\delta P_0&=-\delta\rho\star\Phi
\end{align}
and introducing the quasi-particle energies 
\begin{equation}
\frac{\rho(\theta)}{P(\theta)}=\frac{e^{-\epsilon(\theta)}}{1+e^{-\epsilon(\theta)}}\,,
\qquad\qquad \frac{\bar{\rho}_0(\theta)}{P_0(\theta)}=\frac{e^{-\epsilon_0(\theta)}}{1+e^{-\epsilon_0(\theta)}}
\end{equation}
we arrive at the following TBA equations:
\begin{subequations}
\begin{align}
\epsilon(\theta)&=\frac{\mu}{T}e^{\theta}-(\Phi\star
(L-L_0))(\theta)\,,\\
\epsilon_0(\theta)&=(\Phi\star L)(\theta)\,,
\end{align}
\label{TBAeq}
\end{subequations}
where \(L(\theta)=\log(1+e^{-\epsilon(\theta)})\),
\(L_0(\theta)=\log(1+e^{-\epsilon_0(\theta)})\). \

Now using these equations we can write the extremum of the free
energy:
\begin{equation}
\frac{F}T=-\frac1{2\pi}\int\ud\theta (\mu e^{\theta} R+\Psi(\theta))L(\theta)\,.
\end{equation}
Since the partition function on the cylinder behaves for large \(R\)
as 
\begin{equation}
\log Z_{\alpha\beta}=-\frac{F}T\approx\log(g_\alpha g_\beta)-RE_0^\text{circ}=\log(g_\alpha g_\beta)-\frac{r}l\frac{c_\text{eff}\,\pi}6
\label{scalingZ}
\end{equation}
we obtain
\begin{equation}
\log(g_\alpha g_\beta)=\frac1{2\pi}\int_{-\infty}^\infty\ud\theta\, \Psi(\theta)\log(1+e^{-\epsilon(\theta)})\,.
\end{equation}
This formula is the same given in \cite{leclair} but it is different
from the formula of Ahn and Nepomechie \cite{an} in which
\(L_0(\theta)\) is used instead of \(L(\theta)\). In Figure
\ref{tbaflow} \(\log g_\alpha\) is plotted against \(\log\frac{1}T\).

In the UV limit \(\theta_\text{B}\to-\infty\), the integrand is
non-vanishing if \(\theta\to-\infty\) and similarly, in the IR limit
\(\theta_\text{B}\to\infty\) the non-zero contribution comes from the
\(\theta\to\infty\) domain. Using the fact that
\(\int_{-\infty}^\infty\ud\theta\, \Phi(\theta)=-2\pi\) we find from
the TBA equations \eqref{TBAeq} that
\begin{subequations}
\begin{alignat}{2}
L(-\infty)&=\log(1+\sqrt{2})\,,&\qquad L(\infty)&=0\,,\\
L_0(-\infty)&=\log(2+\sqrt{2})\,,&\qquad L_0(\infty)&=\log2\,.
\end{alignat}
\end{subequations}
Now if one of the boundaries has reflection factor \(\hat R_0\) and
the other one has \(\hat R_I\), then from \eqref{Rs}, \eqref{phidef}
and \eqref{psidef} \(\Psi(\theta)=-\Phi(\theta-\theta_\text{B})\),
so
\begin{equation}
\log\left(\frac{g_\alpha^{\text{UV}}}{g_\alpha^\text{IR}}\right)=\log(1+\sqrt{2})=\log\left(\frac{g_{1,3}}{g_{1,1}}\right)
\end{equation}
since the factor \(g_\beta=g_{1,1}\) cancels. Thus the TBA predicts
the flow \((1,3)\to(1,1)\), which is different from the prediction of
\cite{an}. We will see that TCSA supports our result.

\begin{figure}[h!]
\centering
\includegraphics{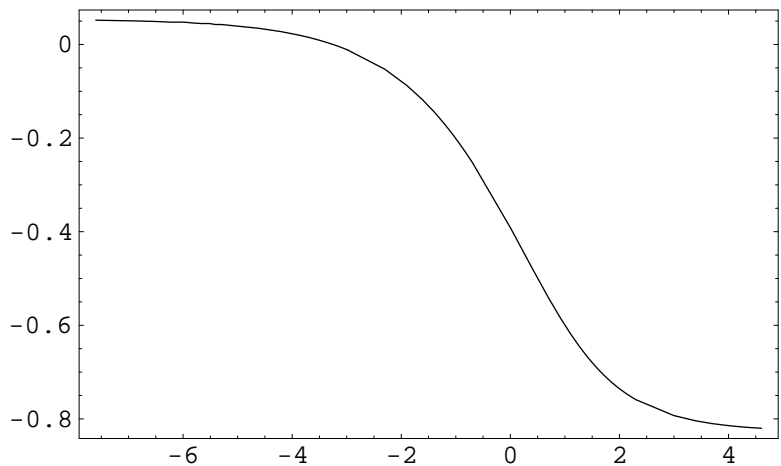}
\caption{TBA: \(\log(g)\) along the flow}
\label{tbaflow}
\end{figure}

\end{section}

\begin{section}{Flows with TCSA}
TCSA can be used to explore the fixed point of the renormalisation
group flows. These flows can be implemented by varying the volume
\(r\) while keeping the coupling constant \(\kappa\) fixed in the
Hamiltonian \eqref{hamilton}. Equivalently -- and we choose this way --
one can keep \(r\) fixed and vary \(\kappa\) on some
interval. Starting from \(\kappa=0\), which is the ultraviolet (UV)
limit, the matrix \(\hat{h}\) can be diagonalised at different values
of \(\kappa\). The flow approaches a fixed point, that is a new
supersymmetric conformal boundary condition. We should observe the
eigenstates rearranging themselves into some degeneracy pattern from
which one can identify the modules (the boundary conditions) using the
characters and weight differences of the supersymmetric minimal model.

It is important that the errors of the TCSA diagonalisation cannot be
controlled easily. For example, it may happen that before the flow
reaches the scaling region the truncation errors start to dominate. If
we use various cuts and find that the flow picture does not change
drastically (only the precision of the result gets higher with higher
cuts), then it means that the unpleasant case mentioned above does not
happen.

Of course one can not establish the endpoint of the renormalisation
group flow using TCSA. What one can see is that the flow goes in the
vicinity of some superconformal boundary condition. The exact infrared
fixed point can never be reached by TCSA because of the truncation. We
are looking for the range where the TCSA trajectory is closest to the
fixed point.

The characters of the (1,3) and \((1,1)\) highest weight
representations are 
\begin{subequations}
\begin{multline}
\chi_{1,3}(q)=1 + q^{1/2} + q +q^{3/2} + 2 q^2 +2 q^{5/2} + 2 q^3 + 3 q^{7/2} + 4 q^4 + 5 q^{9/2} + 5 q^5\\ 
+6 q^{11/2} + 8 q^{6} + 9 q^{13/2} + 10q^7 + \dots
\end{multline}
\vspace{-1.2cm}
\begin{multline}
\chi_{1,1}(q)=1 \qquad\qquad\;\;+q^{3/2} + q^2 + q^{5/2} +q^3 + q^{7/2} + 2q^4 + 2q^{9/2} + 2q^5 \\ 
+3q^{11/2} + 4q^6 + 4q^{13/2} + 4q^7 + \dots
\end{multline}
\end{subequations}

\begin{figure}[h!]
\centering
\subfigure[Integer levels]{\resizebox{73mm}{!}{\includegraphics{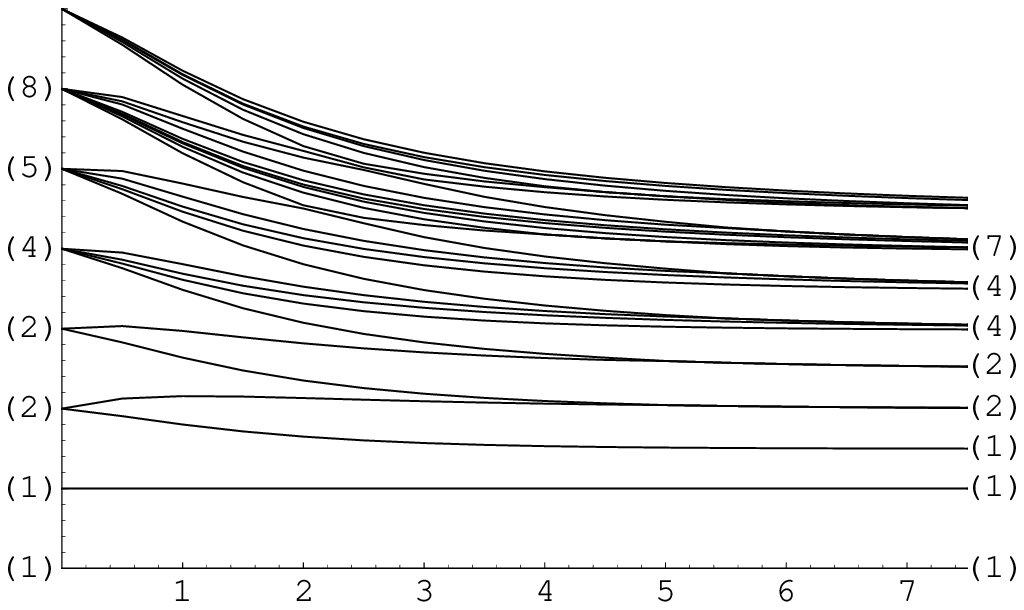}}\label{intflow}}
\ \ 
\subfigure[Half-integer levels]{\resizebox{73mm}{!}{\includegraphics{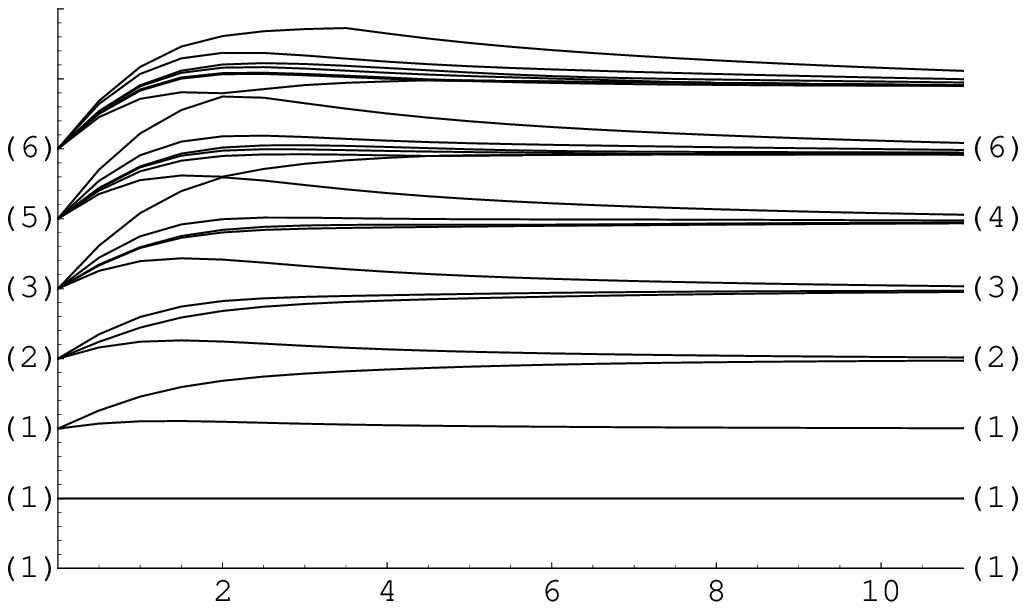}}\label{halfintflow}}
\caption{The RG flows starting from the (1,3) NS and \(\widetilde{\text{NS}}\) b.\ c.-s in SM(2,8)}
\label{28flows}
\end{figure}

In the TCSA calculations the dimension of the Hilbert space was 393
for the integer levels and 344 for the half-integer ones. In Fig.\
\ref{28flows} the normalised energy differences,
\((\varepsilon_i-\varepsilon_0)/(\varepsilon_1-\varepsilon_0)\) are
plotted, and the degeneracies are shown in parentheses. We have found
that starting with the integer (half-integer) levels of the module
(1,3) the flow tends to a degeneracy pattern corresponding to the
integer (half-integer) levels of the module (1,1). Since we did not
perturb the (1,1) boundary we can assume that this boundary condition
remained the same during the flow. Taking into account that (1,1) is
bosonic and (1,3) fermionic, this -- keeping in mind the conclusions
after equation \eqref{hilbert} -- leads to the conclusion that the
flows are
\begin{subequations}
\begin{align}
(1,3)_\text{NS}&\longrightarrow(1,1)_{\widetilde{\text{NS}}}\,,\\
(1,3)_{\widetilde{\text{NS}}}&\longrightarrow(1,1)_\text{NS}
\end{align}
\end{subequations}
in perfect agreement with the results of the TBA analysis based on our
proposed reflection factor.

The flow of the boundary entropy, \(\log(g)\) can also be explored
numerically using TCSA \cite{drtw}. The trace in the partition
function \eqref{hilbert} can be approximated by the finite sum over
the TCSA eigenvalues. For each value of \(l\) one must find the
scaling region, the domain in \(r\) for which \(\log Z\) behaves as in
\eqref{scalingZ}. The volume \(r\) should be great enough for this
scaling behaviour, but for too large values the truncation errors
start to dominate and spoil this simple linear form. In Figure
\ref{dlogzplot} it can be seen that the scaling region is centered
around \(r/l=3\) and in Figure \ref{logzplot} the logarithm of the
partition function is plotted.

The partition function will depend only on the combination
\(x=hL^{1/2-\Delta_{1,3}}=hL^{3/4}=\kappa l^{3/4}\). After making the
linear fit in \(r/l\) along the scaling region for different values of
\(x\) one gets the product of the \(g\)-functions as a function of
\(x\). However this function generally contains a linear term in \(L\)
which corresponds to the free energy density coming from the
boundaries. In order to get the correct final result this term should
be subtracted. We extracted this term numerically from the large \(L\)
behaviour of the naive fits for the \(g\)-function. 

The numeric results are compared with the TBA data in Table
\ref{gtabl} and they are plotted in Figure \ref{tbatcsaflow}
together. Unfortunately there is no systematic method for determining
the TCSA errors, but one can make estimates for them. One way is
comparing the results obtained at different cuts, i.\ e.\ at different
dimensions of the Hilbert space, here we found that for \(x<1\) the
relative error is 0.5--2\%. Another source of error is in the choice
of the scaling region, the corresponding error is 0.2--0.7\%. Finally,
the error of the linear fit itself is about 0.2--0.4\%. So our
estimate for the relative error in the TCSA results is about 1--3\%
and the TBA and TCSA values for the \(g\)-function agree within this
error. Apart from the difference due to the truncation errors in
\(g_\text{TCSA}\) at large values of \(x\) (as in \cite{drtw}) there
is an excellent agreement between the two approaches.

\begin{figure}[t]
\centering
\subfigure[\(\ud(\log Z)/\ud (r/l)\)]{\resizebox{74mm}{!}{\includegraphics{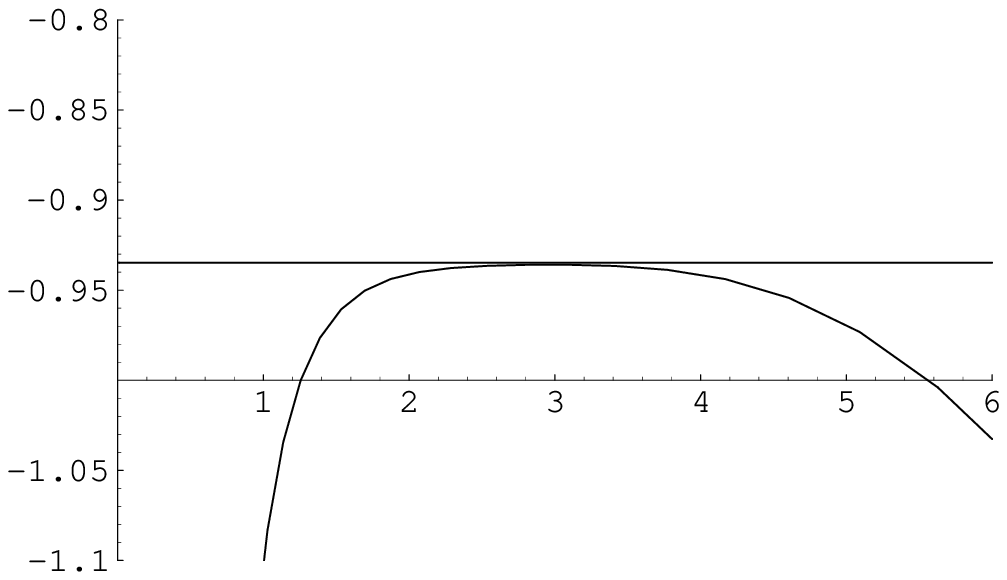}}\label{dlogzplot}}
\subfigure[\(\log Z\) vs. \(r/l\)]{\resizebox{74mm}{!}{\includegraphics{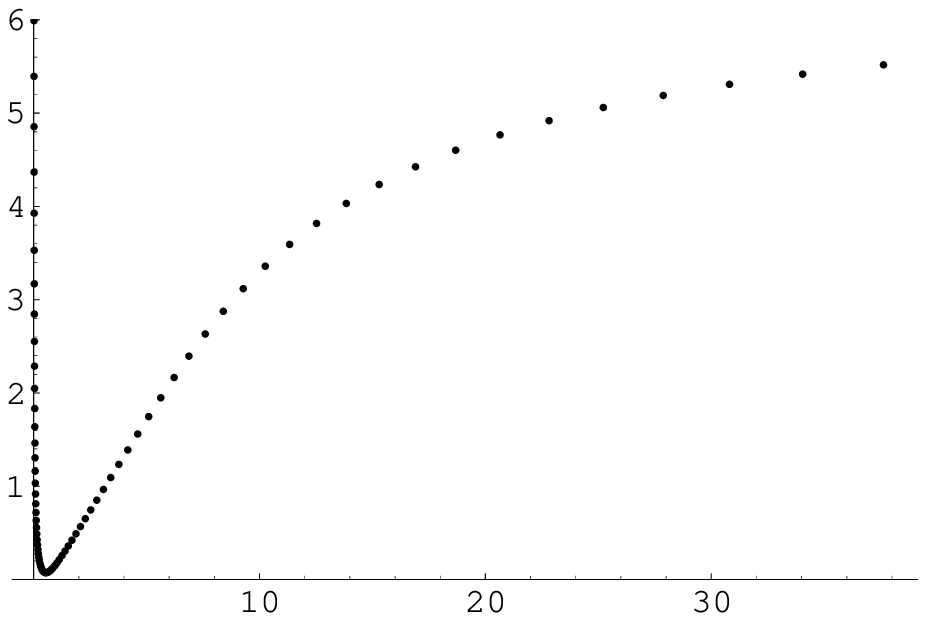}}\label{logzplot}}
\caption{The partition function (\(x=0.1\))}
\label{logzpl}
\end{figure}

\begin{table}[h!]
\begin{center}
\begin{tabular}{|c|c|c|}
\hline
\(\log x\)&\(\log g_\text{TBA}\)&\(\log g_\text{TCSA}\)\\
\hline
\hline
-4.605  & -0.0900  & -0.0932\\
\hline
-3.912   & -0.1013  & -0.1059\\
\hline
-3.507   & -0.1122  & -0.1203\\
\hline
-2.996   & -0.1337  & -0.1449\\
\hline
-2.302  & -0.1846   & -0.2008\\
\hline
-1.609  & -0.2768   & -0.2941\\
\hline
-0.916   & -0.4264  & -0.4322\\
\hline
-0.693  & -0.4864   & -0.4906\\
\hline
-0.224  & -0.6215   & -0.6177\\
\hline
0      & -0.6837   & -0.6922\\
\hline
0.405   & -0.7817   & -0.7457\\
\hline
0.693   & -0.8347   & -0.9015\\
\hline
0.916   & -0.8659   & -0.9377\\
\hline
1.504   & -0.9169   & -1.2646\\
\hline
\end{tabular}
\caption{TBA and TCSA results for \(\log g\)}
\label{gtabl}
\end{center}
\end{table}

\begin{figure}[h!]
\centering
\includegraphics{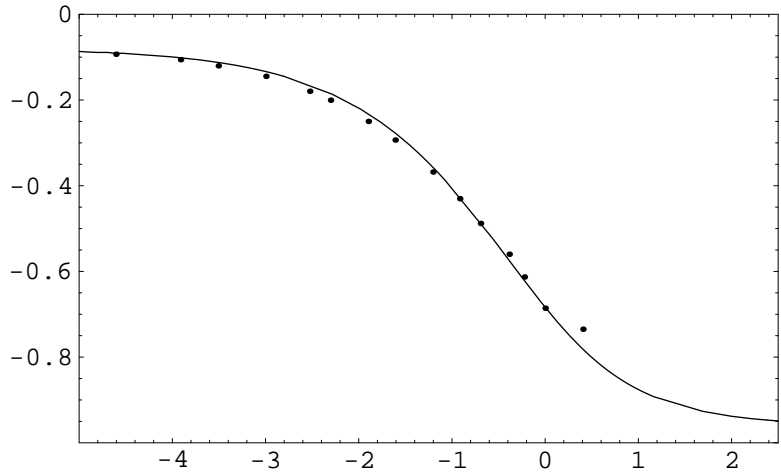}
\caption{\(\log(g)\) vs. \(\log(x)\) along the flow, TCSA (dots) and TBA
  (solid line) results}
\label{tbatcsaflow}
\end{figure}

\end{section}

\newpage

\begin{section}{Generalisation to SM\((2,4n+4)\)}
The supersymmetric Lee--Yang model is the first member of the series
of superconformal minimal models with \(p=2\), the models
SM(2,4n+4). Using TCSA we have found that for \(\kappa>0\) starting
from the even (odd) levels of any NS module we end up at the even
(odd) levels of the (1,1) module (see Figures \ref{21251flows},
\ref{21271flows}). So the IR fixed point seems to be always the (1,1)
boundary condition and depending on the fermion parity of the field
corresponding to the UV boundary condition the flow is of type NS\(\to\)NS
(\(\widetilde{\text{NS}}\to\widetilde{\text{NS}}\)) or
NS\(\to\widetilde{\text{NS}}\) (\(\widetilde{\text{NS}}\to\)NS):
\begin{subequations}
\begin{align}
(1,3)_\text{NS}&\longrightarrow(1,1)_{\widetilde{\text{NS}}}\,,\\
(1,3)_{\widetilde{\text{NS}}}&\longrightarrow(1,1)_\text{NS}\,,
\end{align}
\end{subequations}
\vspace{-0.9cm}
\begin{subequations}
\begin{align}
(1,5)_\text{NS}&\longrightarrow(1,1)_\text{NS}\,,\\
(1,5)_{\widetilde{\text{NS}}}&\longrightarrow(1,1)_{\widetilde{\text{NS}}}\,,
\end{align}
\end{subequations}
\vspace{-0.9cm}
\begin{subequations}
\begin{align}
(1,7)_\text{NS}&\longrightarrow(1,1)_{\widetilde{\text{NS}}}\,,\\
(1,7)_{\widetilde{\text{NS}}}&\longrightarrow(1,1)_\text{NS}
\end{align}
\end{subequations}

\vspace{-0.5cm}

\centerline{\vdots}

\vspace{1.5cm}

\begin{figure}[h!]
\centering
\subfigure[Even levels]{\resizebox{74mm}{!}{\includegraphics{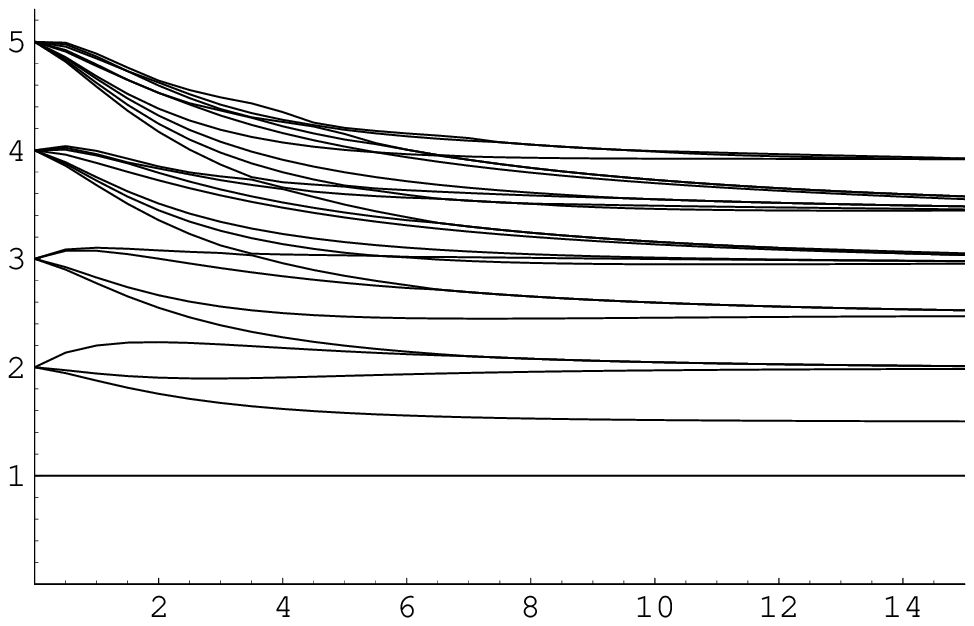}}}
\subfigure[Odd levels]{\resizebox{74mm}{!}{\includegraphics{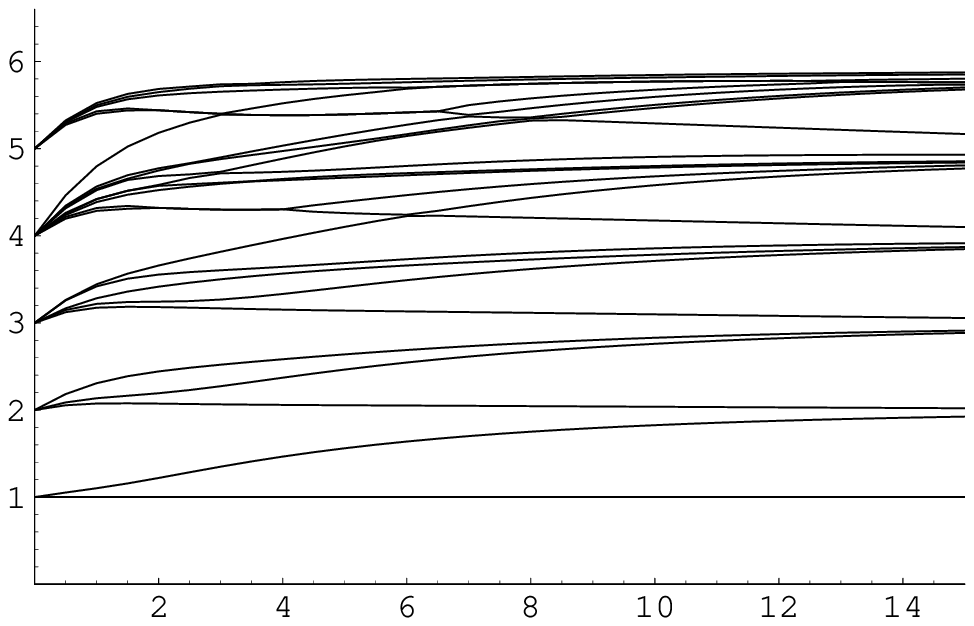}}}
\caption{The RG flows starting from b.\ c.\ (1,5) in SM(2,12)}
\label{21251flows}
\end{figure}

\begin{figure}[h!]
\centering
\subfigure[Even levels]{\resizebox{74mm}{!}{\includegraphics{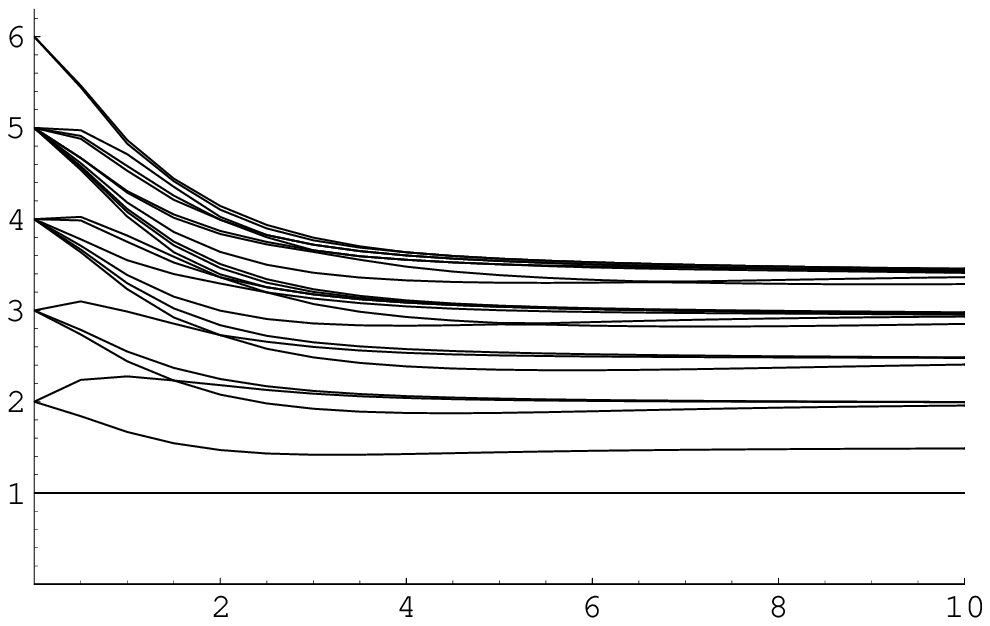}}}
\subfigure[Odd levels]{\resizebox{74mm}{!}{\includegraphics{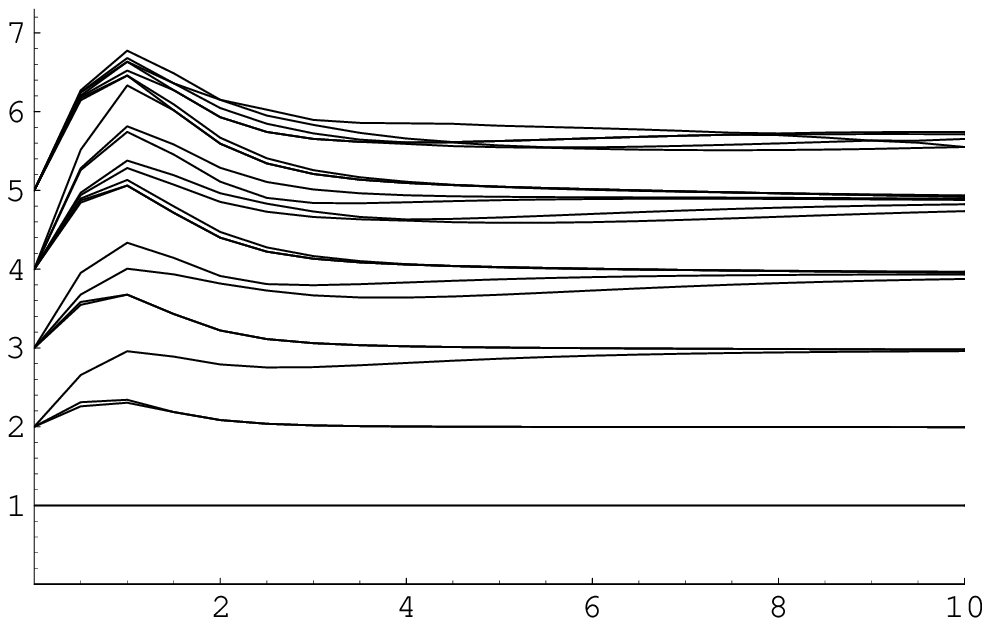}}}
\caption{The RG flows starting from b.\ c.\ (1,7) in SM(2,16)}
\label{21271flows}
\end{figure}

\end{section}

\newpage

\begin{section}{Conclusions}
In this paper we studied the supersymmetric Lee\,--Yang model (SM(2,8)
or M(3,8)) in the presence of boundaries, on a cylinder. First we
proposed reflection factors for the NS type boundary conditions by
considering our model as a reduction of the supersymmetric
sine--Gordon model. The reflection factors of the SSG model contain
two boundary parameters but a consistent reduction requires a
functional relation between them, leaving only one boundary parameter
which is related to the coupling constant of the boundary perturbation
of the SLY model. In the massless limit even this boundary parameter
gets eliminated.

After determining the massless limit of the reflection factors we
compared the energy levels predicted by the Bethe--Yang equation with
the numerical spectrum calculated with the Truncated Conformal Space
Approach. We have found very good agreement for the even levels of the
super Verma module. However, we could not find any reflection factor
that could describe the behaviour of the odd energy levels.

Then we turned to the question of the boundary renormalisation group
flows of the SLY model. The (1,3) boundary was perturbed by the field
\(\hat G_{-1/2}\phi_{1,3}\) which is a relevant, integrable
perturbation that preserves supersymmetry. Using our reflection factor
we wrote down the boundary TBA equations and determined the flow of
the \(g\)-function or equivalently, the boundary entropy. Comparing
the UV and IR values of the \(g\)-function we concluded that the
boundary condition (1,3) flows to the (1,1) boundary condition.

We checked this result using TCSA, diagonalising the Hamilton-operator
of the system at various boundary couplings. From the state content
and degeneracy pattern of the Hilbert space we could identify the
final boundary condition as the boundary condition (1,1), in perfect
agreement with the prediction of the TBA analysis. Finally we
calculated the \(g\)-function along the flow with TCSA, by
approximating the partition function and extracting the \(g\)-function
from the scaling region. The TBA and TCSA results for the change of
the boundary entropy along the flow are in excellent agreement. 

The proposed reflection factors are different from those of Ahn and
Nepomechie \cite{an} but the energy spectrum, the fixed point of the
boundary renormalisation group flow and the TBA \(g\)-function based
on them are in very good agreement with the TCSA results.

At the end of the paper we examined the boundary flows in the
generalisations of the SLY model, the superconformal minimal models
SM\((2,4n+4)\) with TCSA. We found that every Neveu--Schwarz (NS and
\(\widetilde{\text{NS}}\)) boundary condition flows to the (1,1)
boundary condition.

\end{section}

\vspace{-0.2cm}
\subsection*{Acknowledgement}
I would like to thank my supervisor, G\'abor Tak\'acs for his
guidance, help and all the discussions. I am also grateful to Rafael
Nepomechie for the useful discussion on the reflection factor.

\appendix 

\begin{section}{The SSG S-matrix}
\label{ssgs}
The \(n\)th breather in the supersymmetric sine--Gordon model has mass
\begin{equation}
m_n=2m\sin\left(\frac{n\pi}{2\lambda}\right)\,.
\end{equation}
The SSG breather S-matrix can be written in the form
\begin{equation}
S_\text{SSG}^{(i,j)}(\theta)=S_\text{SG}^{(i,j)}(\theta)S_\text{SUSY}^{(i,j)}(\theta)\,.
\end{equation}
%
%
Here \cite{ahn90}
\begin{multline}
S_\text{SG}^{(i,j)}(\theta)=\frac{\sinh(\theta)+i\sin(\frac{i+j}{2\lambda}\pi)}{\sinh(\theta)-i\sin(\frac{i+j}{2\lambda}\pi)}
  \frac{\sinh(\theta)+i\sin(\frac{i-j}{2\lambda}\pi)}{\sinh(\theta)-i\sin(\frac{i-j}{2\lambda}\pi)}\times\\
\prod_{k=1}^{j-1}\frac{\sin^2(\frac{i-j-2k}{4\lambda}\pi+i\frac\theta2)}{\sin^2(\frac{i-j-2k}{4\lambda}\pi-i\frac\theta2)}\frac{\cos^2(\frac{i+j-2k}{4\lambda}\pi+i\frac\theta2)}{\cos^2(\frac{i+j-2k}{4\lambda}\pi-i\frac\theta2)}
\end{multline}
is the sine--Gordon S-matrix and the supersymmetric factor is
\begin{equation}
S_\text{SUSY}^{(i,j)}(\theta)=M^{(i,j)}(\theta)G^{(i,j)}(\theta)
\end{equation}
with
\begin{multline}
M^{(n,m)}(\theta)=\\
\begin{pmatrix}
1+i\,\frac{\sin(\frac{n\pi}{2\lambda})+\sin(\frac{m\pi}{2\lambda})}{\sinh(\theta)}&0&0
&\frac{\sqrt{\sin(\frac{n\pi}{2\lambda})\sin(\frac{m\pi}{2\lambda})}}{\cosh(\frac\theta2)}\\
0&1-i\,\frac{\sin(\frac{n\pi}{2\lambda})-\sin(\frac{m\pi}{2\lambda})}{\sinh(\theta)}&i\,\frac{\sqrt{\sin(\frac{n\pi}{2\lambda})\sin(\frac{m\pi}{2\lambda})}}{\sinh(\frac\theta2)}&0\\
0&i\,\frac{\sqrt{\sin(\frac{n\pi}{2\lambda})\sin(\frac{m\pi}{2\lambda})}}{\sinh(\frac\theta2)}
&1+i\,\frac{\sin(\frac{n\pi}{2\lambda})-\sin(\frac{m\pi}{2\lambda})}{\sinh(\theta)}&0\\
\frac{\sqrt{\sin(\frac{n\pi}{2\lambda})\sin(\frac{m\pi}{2\lambda})}}{\cosh(\frac\theta2)}&0&0&-1+i\,\frac{\sin(\frac{n\pi}{2\lambda})+\sin(\frac{m\pi}{2\lambda})}{\sinh(\theta)}\\
\end{pmatrix}
\end{multline}
and
\begin{gather}
G^{(n,m)}(\theta)=\frac{g(\frac{n+m}{4\lambda})g(\frac12-\frac{n-m}{4\lambda})}{g(\frac12)}\,,\\
g(\Delta)=\frac{\sinh(\frac\theta2)}{\sinh(\frac\theta2)+i\sin(\Delta\pi)}
  \exp\left\{\int_0^\infty\frac{\ud t}t\frac{\sinh(\Delta t)\sinh((1-\Delta)t)}{\cosh^2(\frac{t}2)\cosh(t)}\sinh\left(\frac{i t\theta}{\pi}\right)\right\}\,.
\end{gather}

\end{section}

\begin{section}{SSG reflection factors}
\label{ssgr}
The sine--Gordon reflection factor for the \(n\)th breather is (\cite{ghoshal})
\begin{equation}
R_\text{SG}^{(n)}(\theta)=R_{0}^{(n)}(\theta)R_{1}^{(n)}(\theta)\,,
\end{equation}
where
\begin{multline}
R_{0}^{(n)}(\theta)=(-1)^{n+1}\frac{\cos(\frac{\theta}{2i}+\frac{n\pi}{4\lambda})
  \cos(\frac{\theta}{2i}-\frac\pi4-\frac{n\pi}{4\lambda})\sin(\frac{\theta}{2i}+\frac\pi4)}{
  \cos(\frac{\theta}{2i}-\frac{n\pi}{4\lambda})\cos(\frac{\theta}{2i}+\frac\pi4+\frac{n\pi}{4\lambda})
  \sin(\frac{\theta}{2i}-\frac\pi4)}\\
\times\prod_{l=1}^{n-1}\frac{\sin(\frac\theta{i}+\frac{l\pi}{2\lambda})\cos^{2}(\frac{\theta}{2i}
  -\frac\pi4-\frac{l\pi}{4\lambda})}{
  \sin(\frac\theta{i}-\frac{l\pi}{2\lambda})\cos^{2}(\frac{\theta}{2i}+ \frac\pi4+\frac{l\pi}{4\lambda})}\,.
\end{multline}
\(R_1^{(n)}(\theta)\), which contains the boundary parameters \(\eta\)
and \(\vartheta\) is different depending on whether \(n\) is even or odd:

\begin{equation}
R_{1}^{(2n)}(\theta)=S^{(2n)}(\eta ,\theta)S^{(2n)}(i\vartheta
,\theta)\,,
\end{equation}
where
\begin{equation}
S^{(2n)}(x,\theta)=\prod_{l=1}^{n}\frac{\sin(\frac\theta{i})-\cos(\frac{x}\lambda-(l-\frac12)\frac\pi\lambda)}{\sin(\frac{\theta}{i})+\cos(\frac{x}\lambda-(l-\frac12)\frac\pi\lambda)} \; \frac{\sin(\frac\theta{i})-\cos(\frac{x}\lambda+(l-\frac12)\frac\pi\lambda)}{\sin(\frac\theta{i})+\cos(\frac{x}\lambda+(l-\frac12)\frac\pi\lambda)}
\end{equation}
and
\begin{equation}
R_{1}^{(2n-1)}(\theta)=S^{(2n-1)}(\eta,\theta)S^{(2n-1)}(i\vartheta ,\theta) 
\end{equation}
with
\begin{equation}
S^{(2n-1)}(x,\theta)=\frac{\cos(\frac{x}\lambda)-\sin(\frac\theta{i})}{\cos(\frac{x}\lambda)+\sin(\frac\theta{i})}\prod_{l=1}^{n-1}\frac{\sin(\frac\theta{i})-\cos(\frac{x}\lambda-\frac{l\pi}\lambda)}{\sin(\frac\theta{i})+\cos(\frac{x}\lambda-\frac{l\pi}\lambda)} \; \frac{\sin(\frac\theta{i})-\cos(\frac{x}\lambda+\frac{l\pi}\lambda)}{\sin(\frac\theta{i})+\cos(\frac{x}\lambda+\frac{l\pi}\lambda)}\,.
\end{equation}

In the supersymmetric sine--Gordon model the reflection factors are of
the form (\cite{bssgrevisited,gabor})
\begin{equation}
R^{(n)}_\text{SSG}(\theta)=R^{(n)}_\text{SG}(\theta)\otimes R_\text{SUSY}(\theta)\,.
\end{equation}
In the so-called \(BSSG^+\) case, in which the fermion parity is
conserved during the reflection
\begin{equation}
R_\text{SUSY}(\theta)=
\begin{pmatrix}
\mathcal{A}_+&0\\
0&\mathcal{A}_-
\end{pmatrix}
\label{rsusy}
\end{equation}
with
\begin{equation}
\mathcal{A}_\pm(\theta)=
P\bigx{1.3}(\theta+i\frac\rho2\bigx{1.3})P\bigx{1.3}(\theta-i\frac\rho2\bigx{1.3})\sqrt{2}K(2\theta)2^{-\frac\theta{i\pi}}\,\cos\bigx{1.3}(\frac\theta{2i}\mp\frac\pi4\bigx{1.3})\,,
\end{equation}
where for the \(k\)th breather \(\rho_k=\pi-k\frac\pi\lambda\).
The functions \(K(\theta)\), \(P(\theta)\) are defined as
\begin{align}
K(\theta)&=\frac1{\sqrt{\pi}}\prod_{k=1}^\infty\frac{\Gamma(k-\frac12+\frac\theta{2\pi
    i})\Gamma(k-\frac\theta{2\pi
    i})}{\Gamma(k+\frac12-\frac\theta{2\pi
    i})\Gamma(k+\frac\theta{2\pi i})}\,,\\
P(\theta)&=\prod_{k=1}^\infty\frac{\Gamma^2(k-\frac\theta{2\pi i})}{\Gamma(k-\frac14-\frac\theta{2\pi i})\Gamma(k+\frac14-\frac\theta{2\pi i})}/\{\theta\to-\theta\}\,.
\end{align}
They have the following integral representations:
\begin{align}
P(\theta)&=\exp\left\{-\frac18\int_0^\infty\frac{\ud t}t
\frac{\sinh(\frac{\theta t}{2\pi i})}{\cosh^2(\frac{t}8)\cosh^2(\frac{t}4)}\right\}\,,\\
K(\theta)&=\frac1{\sqrt{\cosh(\frac\theta2)}}\exp\left\{-\frac14\int_0^\infty\frac{\ud t}t
\frac{\sinh(\frac{\theta t}{2\pi i})}{\cosh^2(\frac{t}4)}\right\}\,.
\end{align}
Then \(R_\text{SUSY}(\theta)\) can be written as
\begin{multline}
R_\text{SUSY}(\theta)=\frac{\sqrt{2}2^{-\frac\theta{i\pi}}}{\sqrt{\cosh(\theta)}}
\, \exp\left\{-\frac14\int_0^\infty\frac{\ud t}t
\frac{\cosh(\frac{\rho
    t}{2\pi})+\cosh^2(\frac{t}2)}{\cosh^2(\frac{t}4)\cosh^2(\frac{t}2)}\sinh\left(\frac{\theta t}{i\pi}\right)\right\}\\
\times\begin{pmatrix}
\cos(\frac\theta{2i}-\frac\pi4)&0\\
0&\cos(\frac\theta{2i}+\frac\pi4)
\end{pmatrix}\,.
\end{multline}
\end{section}

\end{document}